\documentclass[jcp,twocolumn,superscriptaddress]{revtex4}
\usepackage{graphicx}
\usepackage{dcolumn}
\usepackage{amsmath}
\usepackage{amssymb}
\usepackage{soul}  
\sethlcolor{yellow}
\usepackage{url}
\usepackage{hyperref}
\usepackage[dvips]{color}
\usepackage{color}
\definecolor{bluegreen}{rgb}{0,0.2,0.8}
\usepackage{subfigure,verbatim,moreverb}
\usepackage{mathrsfs}
\usepackage{longtable}
\usepackage{hhline}

\def\rv{{\bf r}}
\def\uv{{\bf u}}
\def\nn{\nonumber}
\begin{document}
\title{Semilocal exchange hole with an application to range-separation 
density functional}
\author{Jianmin Tao}
\altaffiliation{Corresponding author: jianmin.tao@temple.edu \\
URL: \url{http://www.sas.upenn.edu/~jianmint/}}
\affiliation{Department of Physics, Temple University, Philadelphia,
PA 19122-1801, USA}
\author{Ireneusz W. Bulik}
\affiliation{Department of Chemistry and Department of Physics and
Astronomy, Rice University, Houston, TX 77005}
\author{Gustavo E. Scuseria}
\affiliation{Department of Chemistry and Department of Physics and 
Astronomy, Rice University, Houston, TX 77005}

\date{\today}
\begin{abstract}
Exchange-correlation hole is a central concept in density functional 
theory. It not only provides justification for an exchange-correlation 
energy functional, but also serves as a local ingredient in nonlocal
range-separation density functional. However, due to the nonlocal nature, 
modelig the conventional exact exchange hole presents a great challenge 
to density functional theory. In this work, we propose a semilocal 
exchange hole underlying the Tao-Perdew-Staroverov-Scuseria (TPSS) 
meta-GGA functional. The present model is distinct from previous models at 
small separation between an electron and the hole around the electron. It 
is also different in the way it interpolates between the rapidly varying 
iso-orbital density and the slowly varying density, which is determined by 
the wave vector analysis based on the exactly solvable infinite barrier 
model for jellium surface. Our numerical tests show that the exchange hole 
generated from this model mimics the conventional exact exchange hole quite 
well for atoms. Finally, as a simple application, we apply the hole model 
to construct a TPSS-based range-separation functional. Our tests show that 
this TPSS-based range-separation functional can substantially improve TPSS 
band gaps and barrier heights, without losing much accuracy of molecular 
atomization energies. 

\end{abstract}

\maketitle
\section{Introduction}
Kohn-Sham density functional theory (DFT)~\cite{ks65,Parr89,RMDreizler90} 
is a mainstream electronic structure theory, due to the useful accuracy 
and high computational efficiency. Formally it is exact, but in practice 
the exchange-correlation energy component, which accounts for all 
many-body effects, has to be approximated as a functional of the electron 
density. Development of reliable exchange-correlation energy functionals 
for a wide class of problems has been the central task of DFT. Many 
density functionals have been proposed~\cite{PW86,B88,LYP88,BR89,B3PW91,
B3LYP,PBE96,VSXC98,HCTH,PBE0,HSE03,AE05,MO6L,TPSS03,revTPSS,PBEsol,SCAN15,
Kaup14,Tao-Mo16}, and some of them have achieved remarkable accuracy in 
condensed matter physics or quantum chemistry or both. 

According to the local ingredients, density functionals can be classified 
into two broad categories: semilocal and nonlocal. Semilocal functionals 
make use of the local electron density, density derivatives, and/or the 
orbital kinetic energy density as inputs, such as the LSDA (local 
spin-density approximation)~\cite{VWN80,PW92}, GGA (generalized-gradient 
approximation)~\cite{PBE96,PW91,ZWu06} and meta-GGA~\cite{TPSS03,MO6L,
SCAN15,Tao-Mo16}. Due to the simplicity in theoretical development, 
relatively easy numerical implementation, and cheap computational cost, 
semilocal functionals have been widely-used in electrunic structure 
calculations~\cite{CJCramer09,Quest12,Yangreview,Becke14}. 
Indeed, semilocal DFT can give a quick and often accurate prediction of 
many properties such as atomization energies~\cite{VNS03,PHao13,
LGoerigk10,LGoerigk11}, bond lengths~\cite{CAdamo10,YMo16}, lattice 
constants~\cite{Csonka09,PBlaha09,FTran16}, cohesive energies~\cite{VNS04}, 
etc.

Semilocal DFT has achieved high level of sophistication and practical 
success for many problems in chemistry, physics, and materials science, 
but it encounters difficulty in the prediction of reaction barrier heights, 
band gaps, charge transfer, and excitation energies. A proper description 
of these properties requires electronic nonlocality 
information~\cite{PSTS08}, which is absent in semilocal functionals. 
Nonlocality effect can be accounted for via mixing some amount of exact 
exchange into a semilocal DFT. This leads to the development of 
hybrid~\cite{PBE0,B3PW91,VNS03,JJaramillo03} and range-separation 
functionals~\cite{HSE03,Truhlar11}. The former 
involves the exact exchange energy or energy density, while the latter 
involves the exact as well as approximate semilocal exchange holes. 

There are three general ways to develop the exchange hole: One is from 
paradigm systems such as the slowly varying density~\cite{PW86,PBY96,
Taobook10} and one-electron density~\cite{BR89}, another is from the 
reverse engineering approach~\cite{EP98,tpsshole,Lucian13}, and the third 
is from the density matrix expansion~\cite{Tao-Mo16}. However, semilocal 
exchange holes developed with the reverse engineering approach may not be 
in the gauge of the conventional exchange hole. In the construction of the 
semilocal exchange hole, one must impose the hole to recover the 
underlying semilocal exchange energy density, which is usually not in the 
same gauge of the conventional exchange energy density, due to the 
integration by parts performed in the development of semilocal DFT. 
Examples include the PBE GGA~\cite{EP98} and TPSS 
meta-GGAs~\cite{tpsshole,Lucian13} exchange holes. Many range-separation 
functionals have been proposed~\cite{LKronik11,RBaer10,JCTC16,Kresse11}, 
and some of them have obtained great popularity in electronic structure 
calculations.

Semilocal exchange hole in the gauge of the conventional exchange is of 
special interest in DFT. In this work, we aim to develop an exchange hole, 
which reproduces the TPSS exchange functional. To ensure that our model
hole is in the conventional gauge, we not only impose the exact constraints 
in the conventional gauge (e.g., correct short-range behaviour without 
integration by parts) on the hole model, but also alter the TPSS exchange 
energy density by adding the Tao-Perdew-Staroverov-Scuseria gauge 
function~\cite{tsspg08} with a modification for improving the density tail 
behaviour of the original gauge function. The change in the local exchange 
energy density does not alter the integrated TPSS exchange energy, but it 
largely improves the agreement of the model hole with the conventional 
exact exchange hole. Furthermore, the hole model can generate the exact 
system-averaged exchange hole by replacing the TPSS exchange energy density 
with the gauge-corrected TPSS exchange energy density but with the TPSS
part replaced by the conventional exact exchange energy density. Finally, 
as a simple application, we apply our semilocal exchange hole to construct 
a range-separation functional. This TPSS-based range-separation functional 
yields band gaps of semiconductors and barrier heights in much better 
agreement with experimental values, without losing much accuracy of molecular 
atomization energies.

\section{Conventional exact exchange hole}
For simplicity, we consider a spin-unpolarized density 
($n_\uparrow=n_\downarrow$). For such a density, the exchange (x) energy 
can be written as 
\begin{eqnarray}\label{energy}
E_{\rm x}[n] &=& \int d^3r~ n(\rv) \epsilon_{\rm x}(\rv)
\nn \\ && \nn \\ 
&=& 
\int d^3r~ n(\rv) \frac{1}{2} \int d^3u \frac{\rho_{\rm x}(\rv,\rv+\uv)}{u},
\end{eqnarray}
where $n(\rv)=n_\uparrow+n_\downarrow$ is the total electron density, 
$\epsilon_{\rm x}(\rv)$ is the conventional exchange energy per electron, 
or loosely speaking exchange energy density, and 
$\rho_{\rm x}(\rv,\rv+\uv)$ is the exchange hole at $\rv+\uv$ around an 
electron at $\rv$. It is conventionally defined by 
\begin{eqnarray}\label{xhole}
\rho_{\rm x}(\rv,\rv+\uv) = -\frac{|\gamma_1(\rv,\rv+\uv)|^2}{2n(\rv)}.
\end{eqnarray}
Here $\gamma_1(\rv,\uv)$ is the Kohn-Sham single-particle density matrix 
defined by
\begin{eqnarray}\label{matrix}
\gamma_1(\rv,\uv) = 2\sum_i^{\rm occpu}\phi_i(\rv)^{*}\phi_i(\rv+\uv),
\end{eqnarray}
with $\phi_i(\rv)$ being the occupied Kohn-Sham orbitals. According to the
expression~(\ref{energy}), we can regard the exchange energy as the
electrostatic interaction between reference electrons and their associated 
exchange hole. Therefore, an exchange energy functional cannot be fully 
justified unless the underlying exchange hole has been found.

The exchange hole for a spin-unpolarized density can be generalized to any 
spin polarization with the spin-scaling relation~\cite{OP79}
\begin{eqnarray}\label{spinscaling}
\rho_{\rm x}[n_\uparrow,n_\downarrow] = 
\frac{n_\uparrow}{n}\rho_{\rm x}[2n_\uparrow]+
\frac{n_\downarrow}{n}\rho_{\rm x}[2n_\downarrow].
\end{eqnarray}
Therefore, in the development of the exchange hole, we only need to 
consider a spin-compensated density. Performing the spherical average of 
the exchange hole over the direction of separation vector $\uv$, the 
exchange part of Eq.~(\ref{energy}) may be rewritten as 
\begin{eqnarray}\label{energy2}
E_{\rm x}[n] = \int_0^\infty  du~4\pi u^2 \int d^3r~ n(\rv)
\frac{\langle\rho_{\rm x}(\rv,u)\rangle_{\rm sph}}{2u},
\end{eqnarray}
where $\langle\rho_{\rm x}(\rv,u)\rangle_{\rm sph}$ is the spherical average 
of the exchange hole defined by
\begin{eqnarray}
\langle\rho_{\rm x}(\rv,u)\rangle_{\rm sph} = 
\int \frac{d\Omega_{\uv}}{4\pi}\rho_{\rm x}(\rv,\rv+\uv).
\end{eqnarray}
This suggests that the exchange energy does not depend on the detail of the 
associated hole. Re-arranging Eq.~(\ref{energy2}) leads to a simple 
expression
\begin{eqnarray}\label{energy3}
E_{\rm x}[n] = N
\int du~4\pi u^2 \frac{\langle \rho_{\rm x}(u)\rangle}{2u},
\end{eqnarray}
where $\langle \rho_{\rm xc}(u)\rangle$ is the system average of the 
exchange hole defined by
\begin{eqnarray}\label{ave}
\langle \rho_{\rm x}(u)\rangle = \frac{1}{N}
\int d^3r~ n(\rv)\langle\rho_{\rm x}(\rv,u)\rangle_{\rm sph}.
\end{eqnarray}
Here $N = \int d^3~n(\rv)$ is the number of electrons of a system. 

Although the conventional exchange hole of Eq.~(\ref{xhole}) satisfies the 
sum rule,
\begin{eqnarray}\label{sumrule}
\int d^3u~ \rho_{\rm x}(\rv,\uv) = -1,
\end{eqnarray}
the most important property of the exchange hole, the exchange hole under a 
general coordinate transformation does not. Nevertheless, its system average 
always satisfies the sum rule
\begin{eqnarray}\label{sumrule-system}
\int d^3u~ \langle \rho_{\rm x}(u)\rangle = -1.
\end{eqnarray}
This is a constraint that has been imposed in the development of semilocal
exchange hole. While the exchange energy is uniquely defined, the exchange 
energy density $\epsilon_{\rm x}(\rv)$ and the exchange hole 
$\rho_{\rm x}(\rv,\rv+\uv)$ are not. For example, these local quantities 
can be altered by a general coordinate transformation, or arbitrarily 
adding the Laplacian of the electron density, without changing the exchange
energy~\cite{tsspg08,KBurke98}.

\section{Constraints on the exchange hole}
The conventional exchange hole is related to the pair distribution 
function $g_x(\rv,\rv')$ by
\begin{eqnarray}
n(\rv)\rho_{\rm x}(\rv,\rv')=n(\rv)n(\rv')g_x(\rv,\rv').
\end{eqnarray}
In general, a semilocal exchange hole can be written as 
\begin{eqnarray}\label{Jdefinition}
n(\rv)\rho_{\rm x}(\rv,\rv+u)=n^2(\rv) J_{\rm x}(s,z,u_f),
\end{eqnarray}
where $J(s,z,u_f)$ is the shape function that needs to be constructed, 
with $s=|\nabla n|/(2k_f n)$ being the dimensionless reduced density 
gradient, $k_f=(3\pi^2 n)^{1/3}$ being the Fermi wave vector, 
$z=\tau_W/\tau$, and $u_f=k_f u$. Here $\tau_W = |\nabla n|^2/8n$ is the 
von Weiz\"ascker kinetic energy density, and $\tau$ is the Kohn-Sham 
orbital kinetic energy density given by
\begin{eqnarray}
\tau(\rv)=\sum_{i}^{\rm occup.} |\nabla \phi_{i}(\rv)|^2.
\end{eqnarray}
Here $\phi_{i}$ are the occupied Kohn-Sham orbitals with spin summed over.

\subsection{Constraints on the shape function}
We will seek for a shape function that satisfies the following 
constraints:
\begin{enumerate}
\item[i.] On-top value
\begin{eqnarray}
J(s,z,0) = -1/2.
\end{eqnarray}
\item[ii.] In the uniform-gas limit,
\begin{eqnarray}
J^{\rm unif}(u_f)=
-\frac{9}{2}\bigg[\frac{{\rm sin}(u_f)-{\rm cos}(u_f)}{u_f^3}\bigg].
\end{eqnarray}
\item[iii.] Normalization 
\begin{eqnarray}
\frac{4}{3\pi}\int_0^\infty du_f~u_f^2 J(s,z,u_f) = -1. 
\end{eqnarray}
\item[iv.] Negativity 
\begin{eqnarray}
J(s,z,u_f) \le 0.
\end{eqnarray}
\item[v.] Energy constraint
\begin{eqnarray}\label{energyconstraint}
\frac{8}{9}\int_0^\infty du_f~u_f J(s,z,u_f) = -F_x^{\rm TPSS}(s,z).  
\end{eqnarray}
\item[vi.] Small-$u$ behaviour
\begin{eqnarray}
\lim_{u_f \to 0}\frac{\partial^2 J(s,z,u_f)}{\partial u_f^2} = 
L(s,z),
\end{eqnarray}
where $L(s,z)$ will be discussed below. This is a constraint for the 
conventional exchange hole. That means no integration by parts is performed.
\item[vii.] In the large-gradient limit, the TPSS enhancement factor 
approaches PBE enhancement factor. In this limit, the TPSS shape function 
should also approach the PBE shape function, i.e.,
\begin{eqnarray}
\lim_{s \to \infty}J(s,z,u_f) = J^{\rm PBE}(s,u_f).
\end{eqnarray} 
\end{enumerate}
Among these constraints, (vi) is for the conventional exchange hole, while 
(vii) is a constraint used in the development of TPSS functional. These two
constraints will be discussed in detail below. In previous 
works~\cite{tpsshole,Lucian13}, constraint (vi) was used with integration 
by parts and thus is not for the conventional exchange hole, and (vii) was 
not considered.

\subsection{Small-$u$ behaviour and large-gradient limit}
Expanding the spherically-averaged exchange hole up to second order in 
$u$ yields
\begin{eqnarray}\label{smallu}
\langle\rho_x(\rv,u)\rangle_{\rm sph} &=& 
-\frac{1}{2}n + \frac{1}{12}\bigg[  
 4 \bigg( \tau -\frac{|\nabla n|^2}{8n}\bigg)-\nabla^2 n\bigg]
\nn \\&& \nn \\ &\times& u^2 + \cdots
\end{eqnarray}
Since the Laplacian of the density tends to negative infinity at a nucleus,
the negativity of the exchange hole in the small-$u$ expansion of 
Eq.~(\ref{smallu}) is not ensured. Therefore, we must eliminate it. In 
previous works, the Laplacian of the density was eliminated by integration 
by parts~\cite{tpsshole}. In order to model the conventional exchange hole, 
instead of performing integration by parts, here we eliminate it with the 
second-order gradient expansion of the kinetic energy density,
\begin{eqnarray}\label{slowlyvarying}
\tau \approx \tau^{\rm unif}+|\nabla n|^2/(72n)+\nabla^2 n/6,
\end{eqnarray}
which is exact for slowly varying densities. This technique has been used 
in the development of TPSS~\cite{TPSS03} and other semilocal 
functionals~\cite{PKZB,revTPSS} as well as in the construction of electron 
localization indicator~\cite{TLZR15}.

Substituting Eqs.~(\ref{smallu}) into Eq.~(\ref{Jdefinition}) and 
eliminating the Laplacian $\nabla^2 n$ via~(\ref{slowlyvarying}) yields   
the small-$u$ expansion of the shape function 
\begin{eqnarray}\label{tpss-shape-expan}
J(s,z,u_f) &=& -\frac{1}{2} + \frac{1}{6}\Big(-\frac{3}{10}\frac{\tau}{\tau^{uni}} 
+ \frac{9}{10}-\frac{5}{6}s^2\Big)u_f^2  \nn \\
&+& \cdots 
\end{eqnarray}
leading to
\begin{eqnarray}
L(s,z) = -\frac{1}{3}\Big(\frac{3}{10}\frac{\tau}{\tau^{uni}} - \frac{9}{10}  
+ \frac{5}{6}s^2\Big).
\end{eqnarray}
For one- or two-electron densities, $L(s,z)$ reduces to
\begin{eqnarray}
L(s,z=1) = \frac{3}{2}\Big(\frac{1}{5}  - \frac{8}{27}s^2\Big),
\end{eqnarray}
while for the uniform gas, $L(s=0,z=0) = \frac{1}{5}$. Note that 
$\lim_{s \to 0}L(s,z=1) \neq \lim_{s \to 0}L(s,z=0)$. 

In the large-gradient limit, the TPSS shape function should recover the 
PBE shape function. This requires that $L(s,z)$ must be merged smoothly 
with the PBE small-$u$ behaviour,
\begin{eqnarray}
L^{\rm PBE}(s) = \Big(\frac{1}{5}  - \frac{2}{27}s^2 \Big).
\end{eqnarray}
We can achieve this with
\begin{eqnarray}
L^{\rm TPSS} &=& \frac{1}{2}{\rm erfc}\bigg(\frac{s^2-s_0^2}{ s_0}\bigg) L(s,z) 
\nn \\ && \nn \\ &+&  
\bigg[1 - \frac{1}{2}{\rm erfc}\bigg(\frac{s^2-s_0^2}{s_0}\bigg)\bigg] 
L^{\rm PBE}(s),
\end{eqnarray}
where ${\rm erfc}(x)$ is the complementary error function defined by
\begin{eqnarray}\label{errorf}
{\rm erfc}(x) = 1 - {\rm erf}(x) = \frac{2}{\pi}\int_x^{\infty} dt~e^{-t^2}.
\end{eqnarray}
$s_0 = 6$ is a switching parameter that defines the point at which the 
small-$u$ behaviour smoothly changes from TPSS to PBE. This choice of $s_0$
ensures that the small-$u$ behaviour of our shape function is essentially 
determined by Eq.~(\ref{tpss-shape-expan}), while it merges into the PBE 
shape function in the large-gradient limit.

\section{Shape function for the TPSS exchange hole}
\subsection{TPSS shape function}
The shape function for the TPSS exchange hole is assumed to take 
the following form
\begin{eqnarray}\label{tpssshape}
J^{\rm TPSS}(u_fs,z) & = & 
\bigg[-\frac{9}{4 u_f^4} \bigg( 1- e ^ {-A u_f^2 }\bigg)  \nn \\
                & + & \bigg(\frac{9A}{4u_f^2} + B + C(s,z) u_f^2 +
                G(s,z) u_f^4 \nn \\ 
&+& K(s,z)u_f^6\bigg)e^{-D u_f^2}\bigg]
e^{-H(s,z)u_f^2}, 
\end{eqnarray}
where $A = 0.757211$, $B = -0.106364$, $D =0.609650$ are determined by the
conditions for the uniform electron gas, while the functions $C(s,z)$, 
$G(s,z)$ and $K(s,z)$ are determined by constraints (iii), (v) and (vi). 
They can be analytically expressed in terms of $H(s,z)$ as
\begin{eqnarray}\label{Cequation}
C &=&  \frac{1}{8} \Big( 4L + 3A^3 + 9A^2H
-9AD^2 -18ADH \nn \\
&+& 8B\lambda \Big) \\
\label{Gequation}
G &=& -\frac{63}{8}\lambda^3\Bigg(F_x^{\rm TPSS} + 
A{\rm ln}\Big(\frac{\beta}{\lambda}\Big) 
+ H{\rm ln}\Big(\frac{\beta}{H}\Big)\Bigg) \nn \\ 
&-&\frac{24}{5}\lambda^{\frac{7}{2}}\Bigg(\frac{3A}{\sqrt{H} + 
\sqrt{\beta}} -\sqrt{\pi}\Bigg) 
+ \frac{603}{40}A\lambda^3 \nn \\
&-& \frac{19}{10}B\lambda^2 
-\frac{11}{10}C\lambda  \\
\label{Kequation}
K &=& \frac{8}{35}\lambda^{\frac{9}{2}} \Bigg(\frac{3A}{\sqrt{H} + 
\sqrt{\beta}} 
-\sqrt{\pi}\Bigg) 
- \frac{12}{35}A\lambda^4 \nn \\
&-& \frac{8}{105}B\lambda^3 - \frac{4}{35}C\lambda^2 -\frac{2}{7}G\lambda,
\end{eqnarray}
where $\lambda = D + H(s,z)$ and $\beta = A + H(s,z)$. Following the 
procedure of Constantin, Perdew, and Tao~\cite{tpsshole} for the 
construction of the original TPSS shape function, here we determine 
$s$-dependence of $H(s,z)$ by fitting to the two-electron hydrogen-like 
density, while $z$-dependence is determined by the wave vector analysis 
of the surface energy.
 
\begin{table*}[ht]
\caption{Parameters of the TPSS shape function $H(s,z=1)$ of 
Eq.~(\ref{interpolationH}) and PBE shape function $H(s)$ of 
Eq.~(\ref{PBEH}) determined by a fit to the two-electron exponential 
density.}
\begin{center}
\begin{tabular}{cccccccc|cccccccc}
\hline
\hline
\multicolumn{8}{c}{$H(s,z=1)$ of Eq.~(\ref{interpolationH})} &
\multicolumn{6}{c}{$H(s)$ of Eq.~(\ref{PBEH})} 
\\ \cline{1-8} \cline{10-16}
\multicolumn{1}{c}{$h_0$} &
\multicolumn{1}{c}{$h_1$} &
\multicolumn{1}{c}{$h_2$} &
\multicolumn{1}{c}{$h_3$} &
\multicolumn{1}{c}{$d_0$} &
\multicolumn{1}{c}{$d_1$} &
\multicolumn{1}{c}{$d_2$} &
\multicolumn{1}{c}{$d_3$} && 
\multicolumn{1}{c}{$p_1$} & 
\multicolumn{1}{c}{$p_2$} &
\multicolumn{1}{c}{$p_3$} &
\multicolumn{1}{c}{$p_4$} &
\multicolumn{1}{c}{$p_5$} &
\multicolumn{1}{c}{$p_6$} & \\ \hline
0.0060 &2.8916 &0.7768 &2.0876 &13.695 &$-0.2219$ &4.9917 &0.7972 & &
0.0302&$-0.1035$& 0.1272& 0.1203& 0.4859& 0.1008 \\
\hline
\hline
\end{tabular}
\end{center}
\label{table1}
\end{table*}
\begin{figure}[t]
\includegraphics[width=\columnwidth]{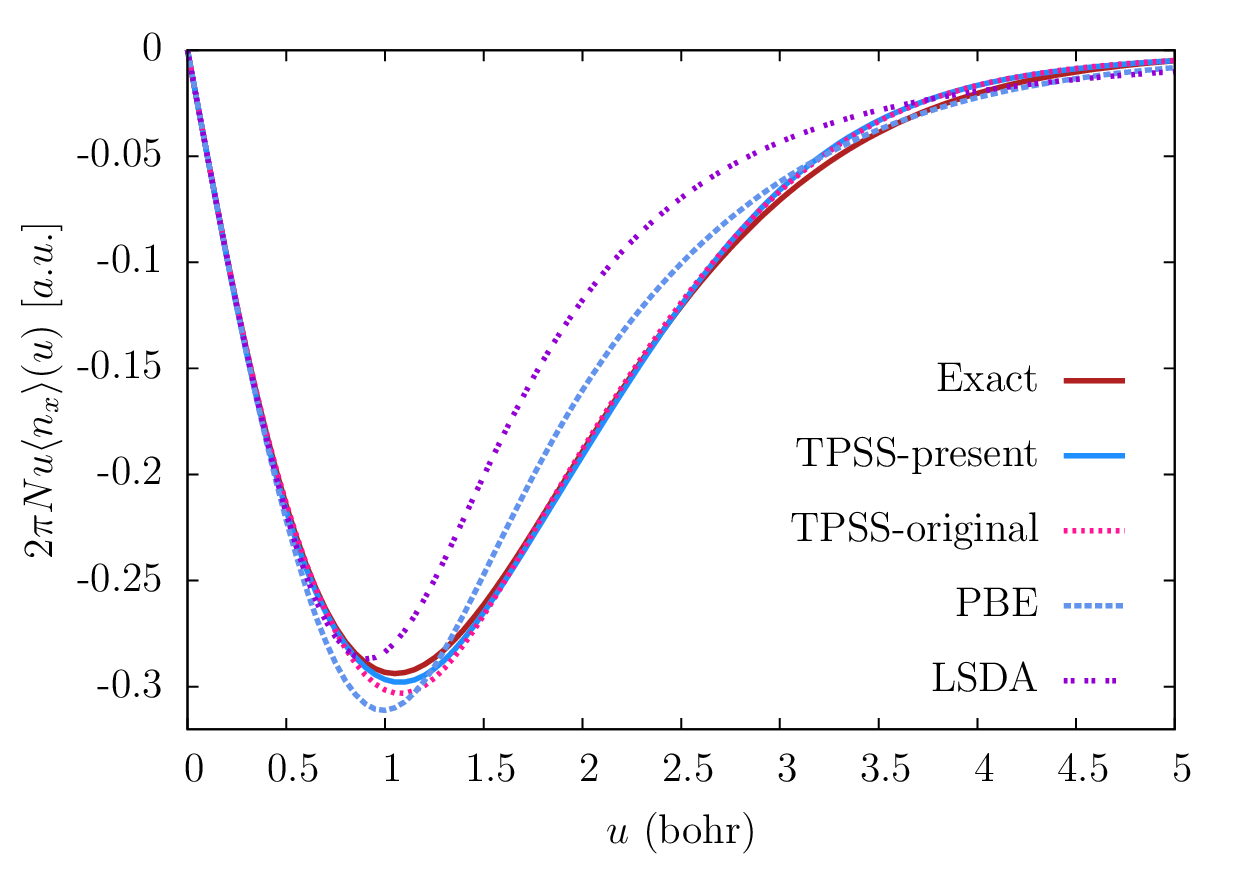}
\caption{System averaged exchange hole for the LSDA, PBE GGA, and TPSS
meta-GGA for the two-electron exponential density. TPSS-original
represents the original TPSS hole model of Constantin, Perdew, and 
Tao~\cite{tpsshole}, while TPSS-present represents the present TPSS hole 
model. The area under the curve is the exchange energy (in hartree):
$E_{\rm x}^{\rm LSDA}=-0.5361$, $E_{\rm x}^{\rm PBE}=-0.6117$,
$E_{\rm x}^{\rm TPSS} = -0.6250$, and $E_{\rm x}^{\rm ex} = -0.6250$. 
Both the original and present TPSS holes yield the same exchange energy, 
due to the same energy constraint.}
\label{figure1}
\end{figure}

\subsection{Gradient dependence of $H(s,z)$}
In so-orbital regions where $z \approx 1$ (e.g., core and density tail 
regions), we assume that the function $H(s,z=1)$ takes the form of
\begin{eqnarray}
H(s,z=1) = 
\frac{h_0 + h_1 s^2 + h_2 s^4 + h_3 s^6}{d_0 + d_1s^2 + d_2 s^4 + d_3 s^6}.
\end{eqnarray}
In the large-gradient regime, the function $H(s,z=1)$ should recover the PBE
$H(s)$ function~\cite{Henderson08}
\begin{eqnarray}\label{PBEH}
H^{\rm PBE}(s) = \frac{p_1s^2 + p_2s^4 +p_3s^6}{1 + p_4s^2 + p_5s^4 + p_6s^6}.
\end{eqnarray}
 For any density between the two regimes, we take the 
interpolation formula,
\begin{eqnarray}\label{interpolationH}
H(s,z=1) &=& \frac{1}{2}{\rm erfc}\bigg(\frac{s^2-s_0^2}{s_0}\bigg)
\mathscr{H}(s,z=1) \nn \\
&+& 
\bigg[1 - \frac{1}{2}{\rm erfc}\bigg(\frac{s^2-s_0^2}{s_0}\bigg)\bigg]
H^{\rm PBE}(s).~~~~~~
\end{eqnarray}
Finally, we insert Eq.~(\ref{interpolationH}) into 
Eqs.~(\ref{Cequation})-(\ref{Kequation}) and perform the fitting procedure 
by minimizing the following quantity 
\begin{eqnarray}
\sum_i u_i \bigg( \big\langle \rho_{\rm x}^{\rm TPSS}(u_i)\big\rangle_{\rm sph}  - 
 \big\langle \rho_{\rm x}^{\rm exact}(u_i) \big\rangle_{\rm sph} \bigg) ^2 
\end{eqnarray}
where $\langle \rho_{\rm x}(u)\rangle_{\rm sph}$ is the system average of 
the exchange hole defined by Eq.~(\ref{ave}). It can be expressed in terms
of the shape function as  
$\langle \rho_{\rm x}(u)\rangle_{\rm sph}  = 
(1/N) \int d^3r n(\rv)^2 J(s,z,u_f)$.
For numerical convenience, we replace the integral with discretsized 
summation. All the parameters for $H(s,z=1)$ and $H(s)$ are listed in 
Table~\ref{table1}.

Figure~\ref{figure1} shows the system-averaged exchange hole 
for the two-electron exponential density evaluated with different hole 
models, compared to the exact one. We can observe from 
Fig.~\ref{figure1} that the present TPSS hole is slightly closer
to the conventional exact hole than the original TPSS hole, but it is much 
closer than the PBE GGA and LSDA holes.

\subsection{Infinite barrier model and wave vector analysis for surface 
energy}
For iso-orbital regions, the $s$-dependence of $H(s,z)$ is determined by 
fitting the model hole to the conventional exact exchange hole for the 
two-electron exponential density. In the uniform-gas limit, our exchange 
hole should correctly reduce to the LDA. This requires that $H(s,z)$ 
vanishes in this limit. To fulfill these considerations, we assume that 
\begin{eqnarray}\label{zdependence}
H(s,z) = H(s,z=1)z^m,
\end{eqnarray}
where $m$ is an integer. In order to determine $m$, we follow the procedure 
of Ref.~\cite{tpsshole} to study the wave-vector analysis (WVA) of 
the surface energy. But instead of using the jellium surface model with 
linearly increasing barrier, here we employ the exactly solvable infinite 
barrier model (IBM). Since the single-particle density matrix and hence the 
electron density of IBM is analytically known, this allows us to obtain 
insight into the $z$-dependence of $H(s,z)$ from this model more easily.

Let us consider a uniform gas of noninteracting electrons subject to the 
infinite potential barrier perpendicular to the $x$ axis ($V \to \infty$ 
for $x < 0$). The one-particle density matrix is given 
by~\cite{LMiglio81,IDMoore76}
\begin{eqnarray}\label{one-particle}
\gamma_1(\rv,\rv') &=& \bar{n}
\bigg[J(u_f)-J\bigg(\sqrt{u_f^2 + 4x_f x_f'}\bigg)\bigg]
\Theta(x) \nn \\ &\times&
\Theta(x')
\end{eqnarray}
where $\Theta(x)$ is a step function, with $\Theta(x)=1$ for $x>0$ and  
$\Theta(x)=0$ for $x\le 0$. $\bar{n}$ is the average bulk valence electron 
density, $x_f=x k_f$, $x_f^{'}=x' k_f$, $u_f=|\rv-\rv'|k_f$, and
\begin{eqnarray}
J(\xi)  =  3 j_1(\xi)/\xi,
\end{eqnarray}
with $j_1(\xi) = {\rm sin}(\xi)/\xi^2-{\rm cos}(\xi)/\xi$ being the 
first-order spherical Bessel function. The electron density can be 
obtained from the single-particle density matrix by taking 
$u=|\rv'-\rv|=0$ in Eq.~(\ref{one-particle}). This yields
\begin{eqnarray}\label{IBMdensity}
n(x) = \bar{n}[1-J(2x_f)]\Theta(x).
\end{eqnarray}

\begin{figure}
\includegraphics[width=\columnwidth]{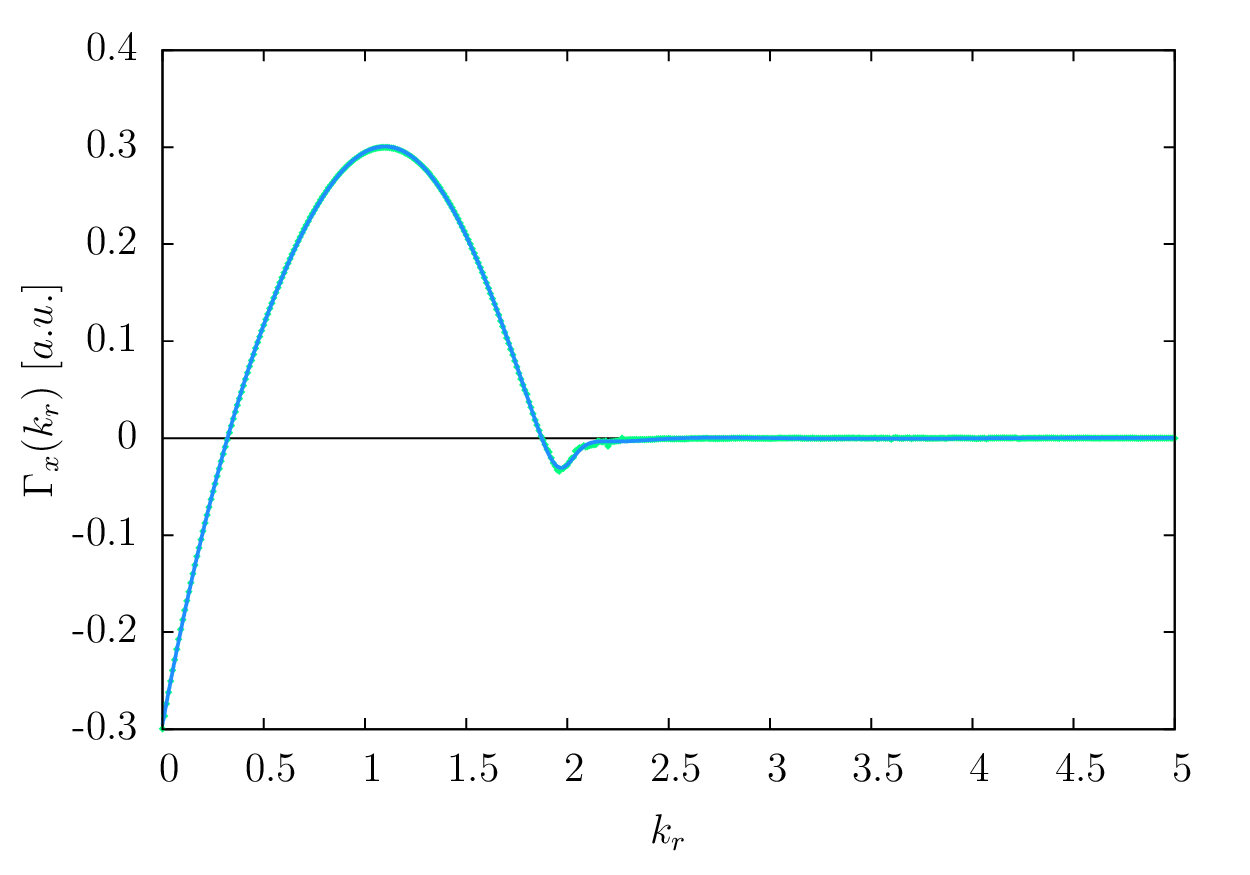}
\caption{$\Gamma(k)$ of Eq.~(\ref{gammak}) and smooth fit.}
\label{figure2}
\end{figure}

The WVA for the surface exchange energy density is given by~\cite{tpsshole}
\begin{eqnarray}\label{gammak1}
\gamma_x(k) = \int_0^\infty du~ 8 k_f u^2 b_x(u) \frac{sin(k u)}{k u} 
\end{eqnarray}
where
\begin{eqnarray}
b_x(u) = \int_{-\infty}^{\infty}dx~ n(x)
\bigg[\rho_x(x,u) -\rho_x^{\rm unif}(u) \bigg].
\label{bx}
\end{eqnarray}
The exchange hole $\rho_x(x,u)$ of IBM can be obtained from the 
one-particle density matrix of Eq.~(\ref{one-particle}). With some algebra,
we can express the WVA surface exchange energy as~\cite{Langreth82} 
\begin{eqnarray}\label{IBMsurface}
\sigma_x = \frac{1}{2}\int_0^\infty dk_r\, \gamma_x(k_r),
\end{eqnarray}
where $k_r = k/k_F$, and $\gamma_x(k_r)$ is given by
\begin{eqnarray}\label{gammak}
\gamma_x(k_r) & = & \frac{8}{k_f^2} \int_0^\infty d u_f\, 
b(u_f)u_f^2 sinc(k_ru_f) \nn \\ &=& \frac{1}{(\pi r_s)^3}\Gamma(k_r),
\end{eqnarray}
and
\begin{eqnarray}
b_x(u_f) =-\frac{\bar{n}^2}{2 k_f} \int_0^\infty dx_f\,i_x(x_f,u_f).
\end{eqnarray}
Here $i_x(x_f,u_f) = \sum_{l = 1 }^6 \chi_l (x_f,u_f)$, with
$\chi_l (x_f,u_f)$ being defined by Eq. (3.18) of Ref.~\cite{Langreth82}.

Figure~\ref{Gamma} shows $\Gamma(k_r)$ of Eq.~(\ref{gammak}). The area under
the curve is proportional to the surface exchange energy. From the electron 
density and density matrix of IBM given by Eqs.~(\ref{one-particle}) 
and~(\ref{IBMdensity}), the exact surface exchange energy can be calculated
with the WVA of Eq.~(\ref{IBMsurface}). Langreth and Perdew~\cite{Langreth82} 
reported that the value of $\sigma_x 10^3 r_s^3$ is $4.0$ a.u., where
$r_s$ is Seitz radius. This value is slightly smaller than the value obtained 
earlier by Harris and Jones~\cite{Harris74} and Ma and Sahni~\cite{CQMa79} 
($4.1$ a.u.). Our present work gives $3.99$ a.u., which might be the most
accurate one.

\subsection{$z$-dependence of $H(s,z)$}
The $z$-dependence of $H(s,z)$ [Eq.~(\ref{zdependence})] can be determined 
by fitting the TPSS hole to the wave vector analysis. We start with the 
specific expressions for the local ingredients of the hole model in IBM. 

From the electron density of Eq.~(\ref{IBMdensity}), the reduced density 
gradient can be explicitly expressed as 
\begin{eqnarray}
s(x_f) = \frac{3}{2 x_f} \frac{|sinc(2 x_f) - J(2x_f)|}{[1-J(2x_f)]^{4/3}}.
\end{eqnarray}
The kinetic energy density can be obtained from the single-particle density 
matrix of Eq.~(\ref{one-particle}). This yields 
\begin{eqnarray}\label{IBMkinetic}
\tau(x_f) &=& k_f^2 \bar{n} \bigg\{ \frac{3}{10} + \frac{1}{2}J(2x_f) + 
\frac{9}{4x_f^2}[sinc(2x_f) \nn \\
&-& J(2x_f)]\bigg\}.
\end{eqnarray}
Finally, the von Weiz\"ascker kinetic energy density can be expressed as
\begin{eqnarray}
\tau_W = \frac{9k_f^2\bar{n}}{8x_f^2}
\Bigg\{\frac{[sinc(2x_f)-J(2x_f)]^2}{1 - J(2x_f)}\Bigg\}.
\end{eqnarray}

\begin{figure}
\begin{center}
\includegraphics[width=\columnwidth]{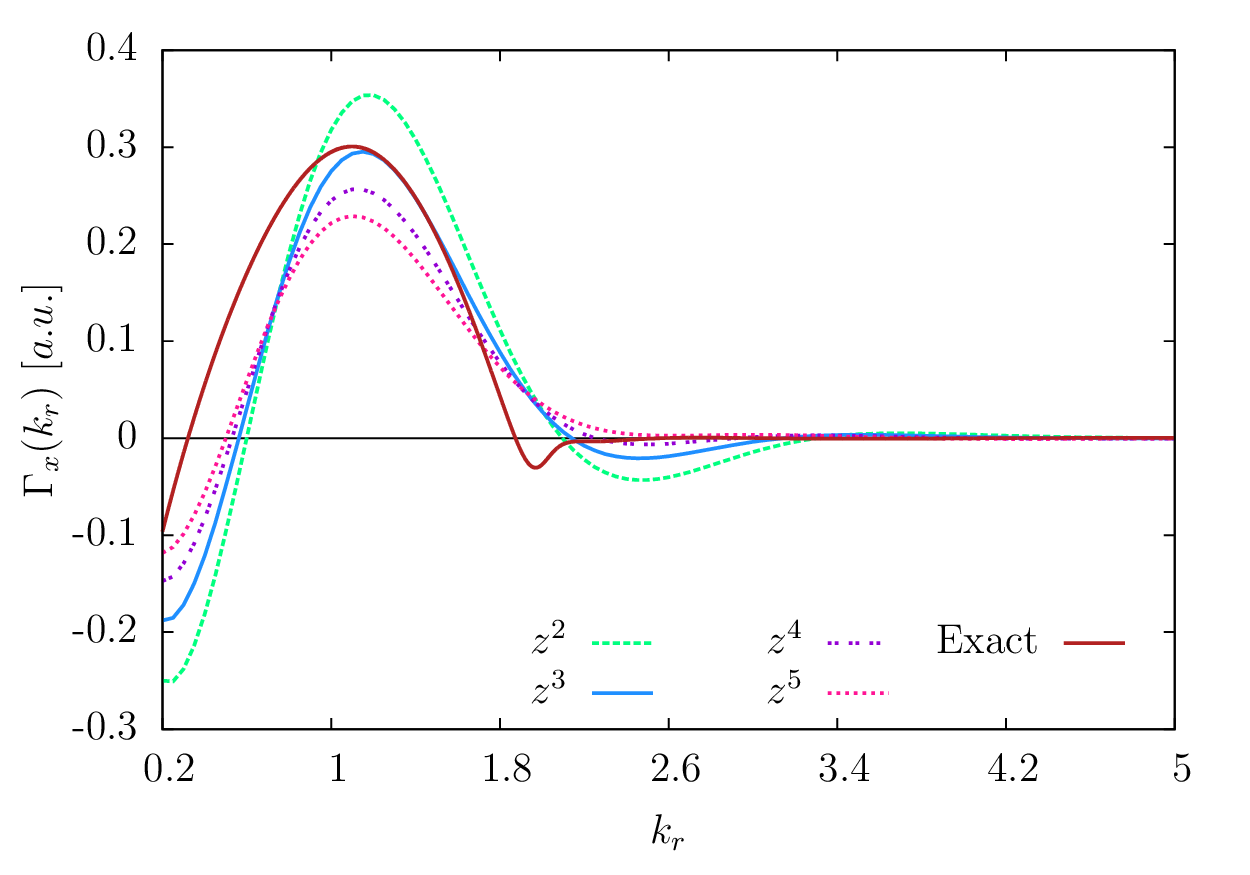}
\caption{Analysis of $z$-dependence of the WVA for the present TPSS hole of
Eq.~(\ref{tpssshape}). $z^3$ curve provides the best fit to the exact one.}
\label{figure3}
\end{center}
\end{figure}

Next, we calculate $\gamma_x$ from the TPSS hole. Starting with $b_x(u)$ 
of Eq.(\ref{bx}), we obtain 
\begin{eqnarray}\label{tpssbx}
b_x(u) &=& \int_0^\infty dx\, n(x)\Big[\rho_{\rm x}^{\rm TPSS}(x,u) -
\rho_{\rm x}^{unif}(u)\Big] \nonumber \\
&=& \frac{\tilde{n}^2}{k_f} \int_0^\infty dx_f~[1-J(2x_f)] \bigg\{
[1-J(2x_f)] \nn \\ &\times&
J^{\rm TPSS}\big(u_f\sqrt[3]{1-J(2x_f)},s(x_f),z(x_f)\big) \nn \\ &-&
J^{\rm unif}(u_f)\bigg\} 
\end{eqnarray}
(Note that $\Theta(x)$ is implicit on the electron density.) Substituting 
Eq.~(\ref{tpssbx}) into Eq.~(\ref{gammak1}), we obtain
\begin{eqnarray}\label{TPSSgamma}
\gamma_x(k) = 
\frac{8 \bar{n}}{3\pi^2} \int_0^\infty dx_f~
\int_0^{\infty} du_f j_x(u_f,x_f,k_r),  
\end{eqnarray}
where 
\begin{eqnarray}
j_x &=& \varrho(x_f)\Big[\varrho(x_f) J^{\rm TPSS}
\big(u_f\sqrt[3]{\varrho(x_f)},s(x_f),z(x_f)\big) \nn \\ &-&
J^{\rm unif}(u_f)\Big]
sinc(k_r u_f) u_f^2
\end{eqnarray}
and $\varrho(x_f)=1 - J(2x_f)$. Rearrangement of Eq.~(\ref{TPSSgamma}) leads
to the final expression
\begin{eqnarray}\label{TPSSWVA}
\gamma_x(k) = \frac{1}{(\pi r_s)^3}\Gamma^{TPSS}(k_r),
\end{eqnarray}
where 
\begin{eqnarray}
\Gamma^{TPSS}(k_r) = 2 \int_0^\infty dx_f\, J_x(x_f,k_r),
\\ && \nn \\
J_x(x_f,k_r) = \int_0^\infty du_f~j_x(u_f,x_f,k_r).
\end{eqnarray}

\begin{figure}
\begin{center}
\includegraphics[width=\columnwidth]{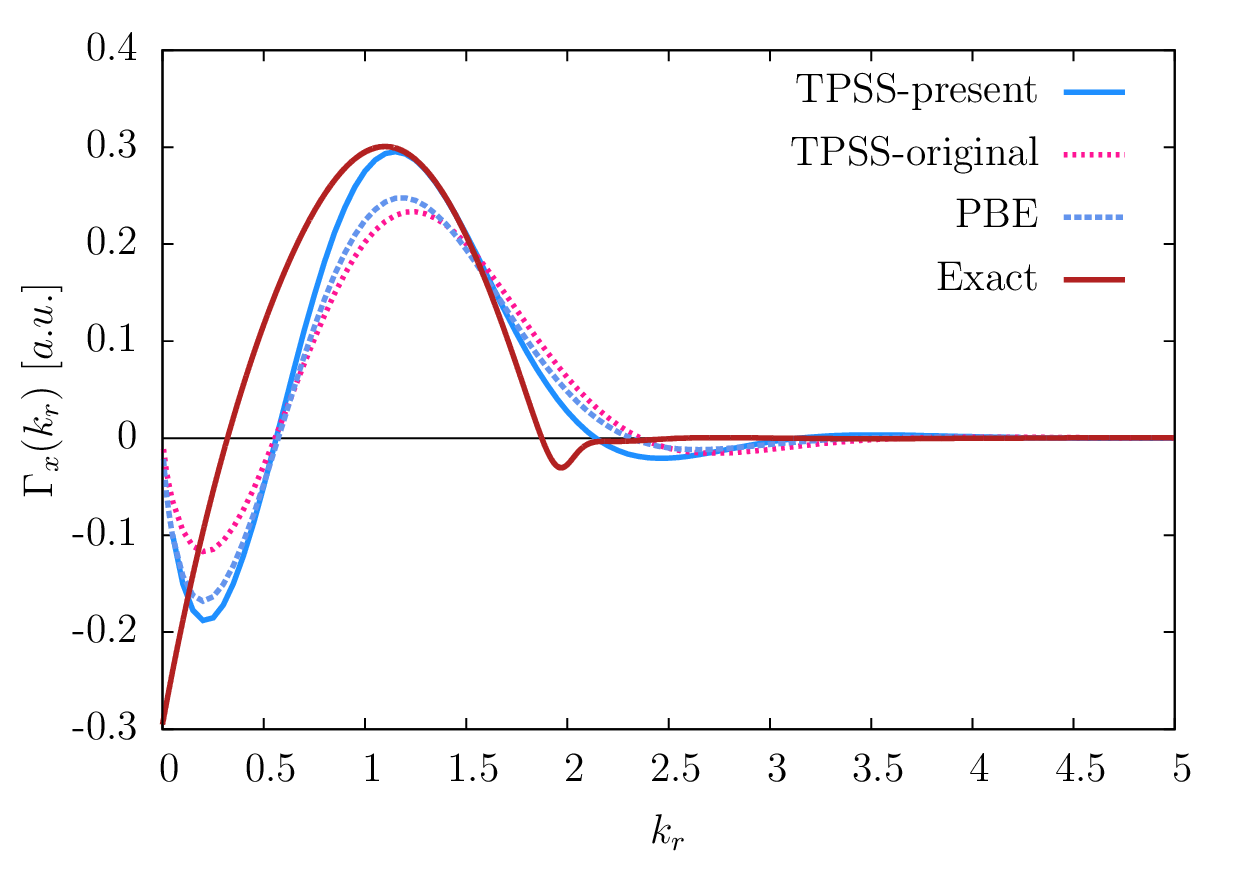}
\caption{Comparison of the WVA for the present and original TPSS hole
models as well as the PBE hole with the exact one. ``TPSS-present'' 
represents the present TPSS hole model, while ``TPSS-original`` represents
the original TPSS model.}
\label{figure4}
\end{center}
\end{figure}

\begin{figure}
\begin{center}
\includegraphics[width=\columnwidth]{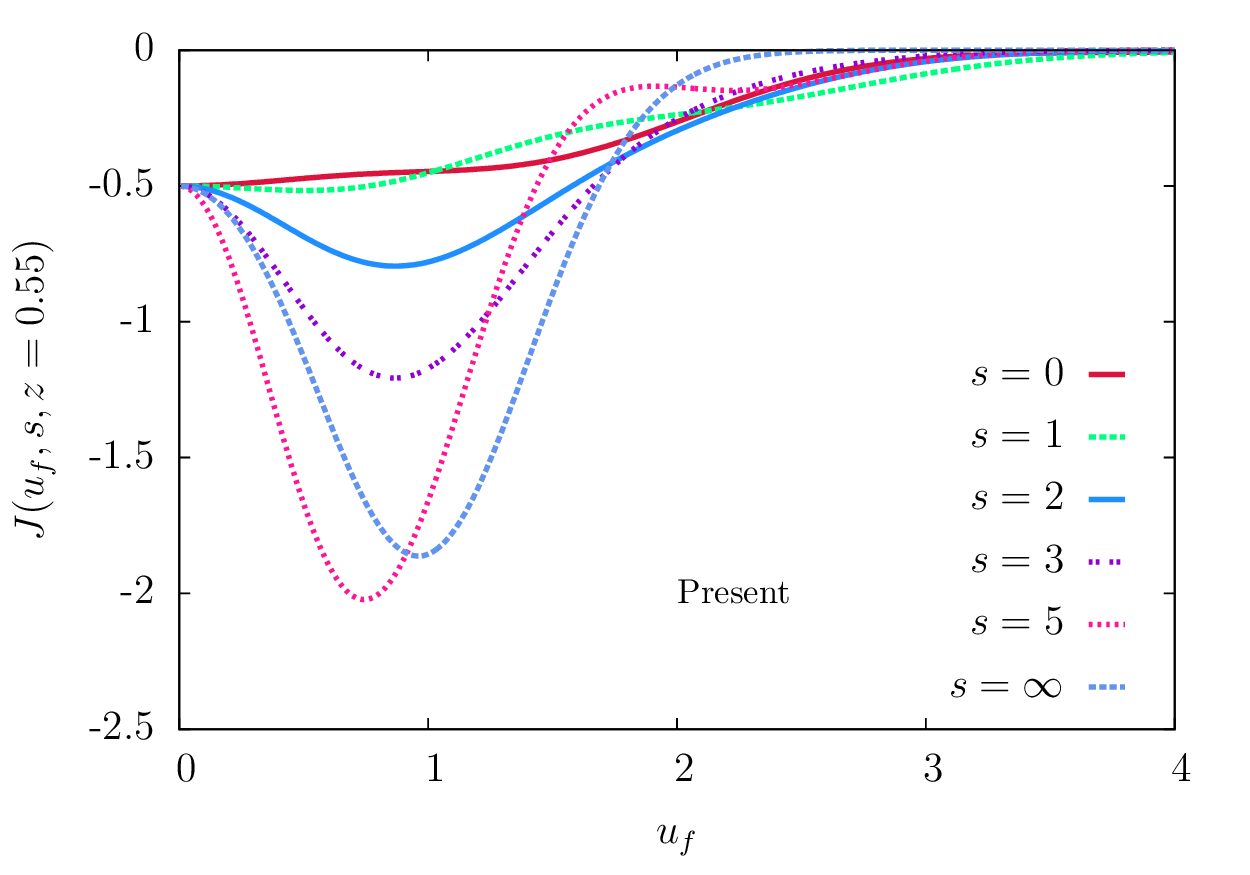}
\caption{Present TPSS shape function of Eq.~(\ref{tpssshape}) for $z=0.55$}
\label{figure5}
\end{center}
\end{figure}

\begin{figure}
\begin{center}
\includegraphics[width=\columnwidth]{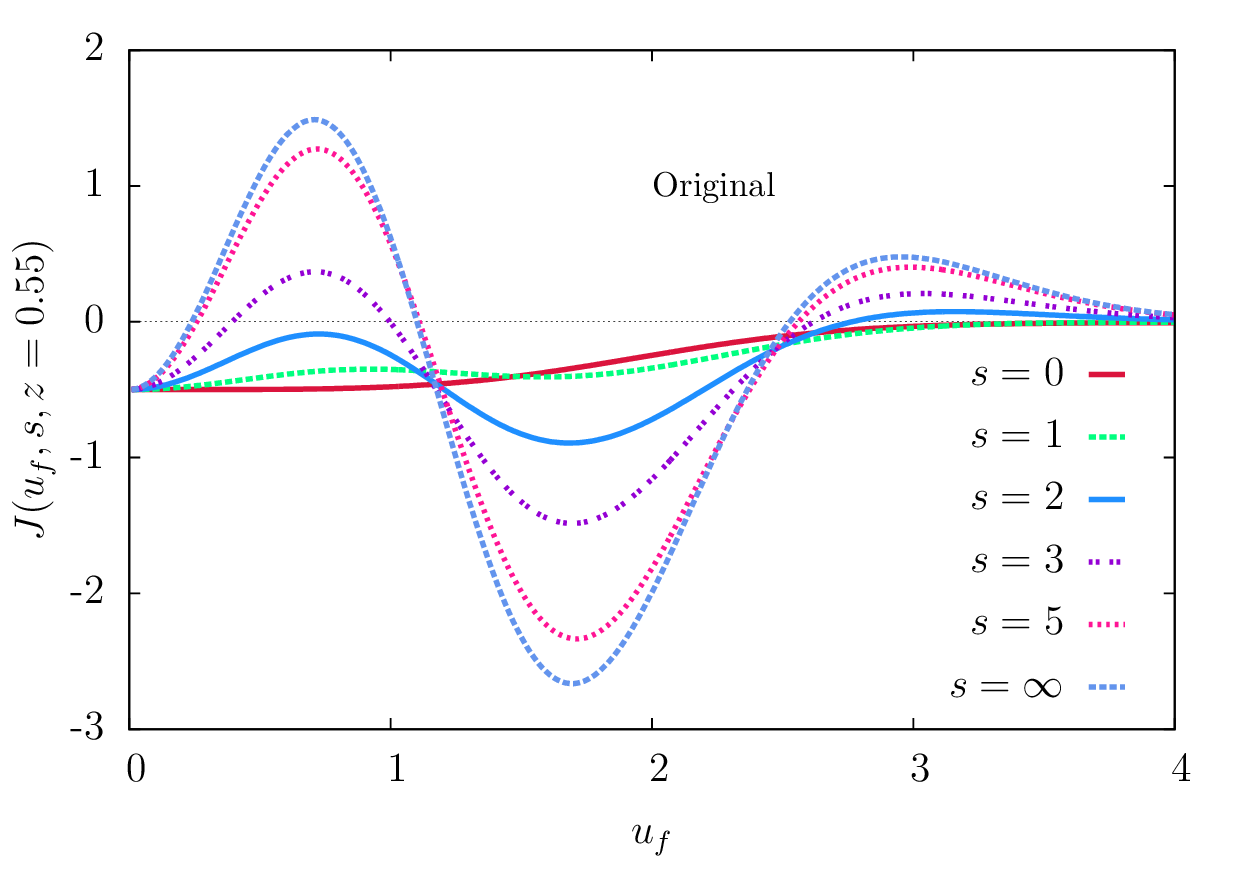}
\caption{Original TPSS shape function for $z=0.55$}
\label{figure6}
\end{center}
\end{figure}

Figure~\ref{figure3} shows the comparison of $H(s,z)$ with different
choices of $m$ to the exact one. From Fig.~\ref{figure3}, we see that
the best choice is $m=3$. Figure~\ref{figure4} shows that, compared to the
LDA, PBE, and original TPSS holes, the WVA of the present model is in 
best agreement with the exact one. To further understand these two models, 
we plot the TPSS shape function of the present and the original models in 
IBM at $z=0.55$, as shown by Figs.~(\ref{figure5}) and~(\ref{figure6}), 
respectively. From Figs.~(\ref{figure5}) and~(\ref{figure6}), we 
observe that while the present model hole is always negative, the original 
TPSS hole can be positive at some values of $u_f$ and $s$.   

To check our wave vector analysis for the surface exchange energy, we 
have computed $\sigma_x$ from 
\begin{eqnarray}
\sigma_x &=& \int_{-\infty}^{\infty} dx\, n(x)\Bigg[\epsilon_x(n) - 
\epsilon_x^{\rm unif} (\bar{n})\Bigg].
\label{AlternativeSigma}
\end{eqnarray}
The results are shown in Table~\ref{tablex}. From Table~\ref{tablex}, we
can see that the surface energy from the WVA of the PBE exchange hole shows
some inconsistency with the surface energy evaluated directly from the 
PBE exchange functional with Eq.~(\ref{AlternativeSigma}), but the surface
energy from the WVA of the TPSS hole (both original and the present version)
agrees very well with the surface energy calculated from TPSS exchange 
functional with Eq.~(\ref{AlternativeSigma}). Furthermore, TPSS surface 
energy is more closer to the exact value than those of the LDA and PBE. 
The LDA significantly overestimates the surface exchange energy. All these 
observations are consistent with those evaluated from jellium surface model.
It is interesting to note that even the original TPSS shape function in 
certain range is positive, the surface energy from the original TPSS hole
is the same as that from the present model. This result is simply due to 
the cancellation of the original hole model between positive values and 
too-negative values at certain $u_f$ values, as seen from the comparison 
of Fig.~(\ref{figure6}) to~(\ref{figure5}). The IBM surface energy 
presents a great challenge to semilocal DFT. It is more difficult to get
it right than the surface energy of jellium model with finite linear 
potential, because the electron density at the surface of IBM is highly
inhomogeneous, due to immediate cutoff at surface, and is too far from the 
slowly varying regime, in which semilocal DFT can be exact (e.g., TPSS 
functional).

\begin{figure}
\begin{center}
\includegraphics[width=\columnwidth]{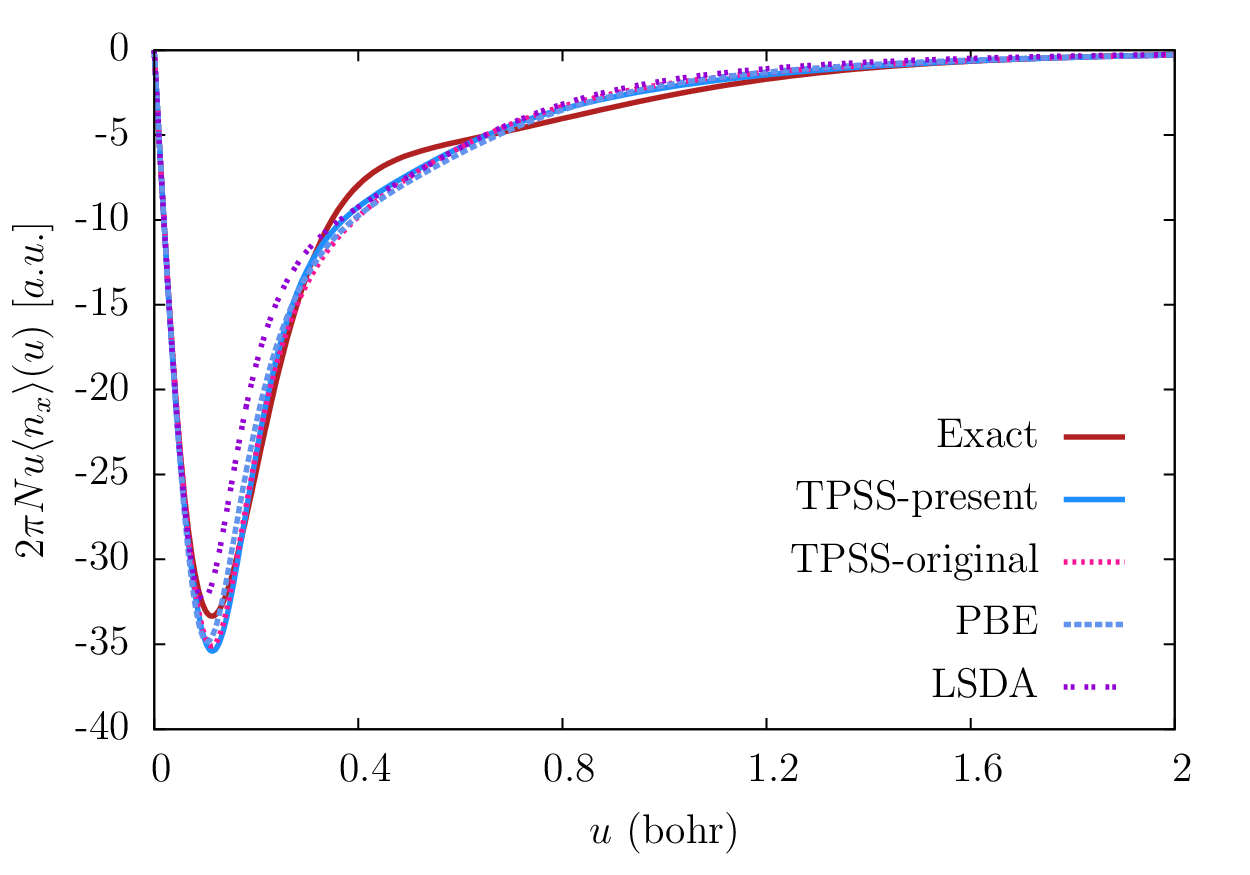}
\caption{Comparison of various system-averaged hole models for the Ne atom.
``TPSS-present'' represents the present TPSS hole model, while
``TPSS-original'' represents the original TPSS model.}
\label{figure7}
\end{center}
\end{figure}

\begin{table}
\caption{Comparison of the surface exchange energies (in a.u.) of IBM 
surface (expressed as $\sigma_x{r_s}^3 {10}^3$) calculated directly with 
exchange energy functionals and with the WVA formula. Exact value 
(obtained in this work) is 3.99 a.u.}
\begin{tabular}{cc|c}
\hline
\hline
        & Eq~.(\ref{AlternativeSigma})  & WVA integration        \\
\hline
LDA     &   6.318                       &   -                    \\
\hline
PBE     &   2.576                       &  2.67                  \\
\hline
TPSS    &   2.945                       &  2.95 (original hole)  \\
        &                               &  2.95 (present hole)   \\
\hline
\hline
\end{tabular}
\label{tablex}
\end{table}

Figure~\ref{figure7} shows the comparison of the LSDA, PBE, and the 
original and present TPSS exchange hole models to the conventional exact 
exchange hole of the Ne atom, in which $z$ is in general different from 0 
(slowly varying density) and 1 (iso-orbital density). The PBE and LSDA 
curves are plotted with the hole models of Ref.~\cite{Henderson08}. 
From Fig.~\ref{figure7} we can see that the present TPSS hole model is 
generally closer to the exact one than the original TPSS hole model, while 
the original TPSS hole model is very close to PBE, but both of them are 
closer to the exact one than the LSDA.

\section{TPSS hole in the gauge of the conventional exact exchange}
The shape function explicitly depends on the enhancement factor via the 
energy constraint of Eq.~(\ref{energyconstraint}). The latter may be 
altered by arbitrarily adding any amount of the Laplacian of the density 
without changing the exchange energy. This ambiguity of the exchange
energy density~\cite{KBurke98} leads to the ambiguity of the semilocal 
exchange hole. Our primary goal of this work is to develop a semilocal 
exchange hole in the gauge of the conventional exact exchange. This is 
partly motivated by the fact that in the development of 
range-separationals, the exact exchange part is usually provided in the 
conventional gauge. 

The exact exchange energy density in the conventional 
gauge can be conveniently evaluated with the Della Sala-G{\"o}rling 
(DSG)~\cite{DSG01} identity resolution  
\begin{eqnarray}
e^x_{conv}(\rv) = 
\frac{1}{2}\sum_{\mu\nu} Q_{\mu\nu}^\sigma \chi_\mu(\rv)\chi_\nu^*(\rv),
\end{eqnarray}
where $Q^\sigma$ is spin block of DSG matrix~\cite{tsspg08}. However, many
semilocal exchange energy densities or enhancement factors of
Eq.~(\ref{energyconstraint}) are not in the gauge of the conventional 
exact exchange, due to the constraints such as the Lieb-Oxford bound and 
the slowly-varying gradient expansion imposed on the enhancement factor. 
For example, for the two-electron exponential density, the conventionally
defined exact enhancement factor is less than 1 near the nucleus, while 
the TPSS enhancement factor is $F_{\rm x}^{\rm TPSS} \ge 1$ by design. To 
construct the TPSS exchange hole in the conventional gauge, we can replace
the original energy density constraint [Eq.~(\ref{energyconstraint})], as 
used in the construction of the original TPSS exchange hole~\cite{tpsshole}, 
with the TPSS exchange energy density or enhancement factor in the 
conventional gauge. In this gauge, the TPSS exchange energy density can be 
written as~\cite{tsspg08}
\begin{eqnarray}
e_{\rm x}^{\rm TPSS,conv}(\rv) = e_{\rm x}^{\rm TPSS}(\rv)-G(\rv),
\end{eqnarray}
where $e_x(\rv)=n(\rv)\epsilon_x(\rv)$. 
Based on the uniform and non-uniform coordinate scaling properties
of the exact exchange energy density, Tao, Staroverov, Scuseria, and 
Perdew (TSSP)~\cite{tsspg08} proposed a gauge function 
\begin{eqnarray}
G(\rv) &=& a\nabla\cdot [f(\rv)\nabla g(\rv)],
\label{Gauge-general} \\ && \nn \\
g &=& \nabla \tilde{\epsilon} \\
\label{ffunction}
f &=& \frac{n/\tilde{\epsilon}^2} 
{1 + c(n/\tilde{\epsilon}^3)^2} 
\bigg(\frac{\tau^W}{\tau}\bigg)^b.
\end{eqnarray}
Here $a = 0.015$ and $c=0.04$ are determined by a fit to the conventional 
exact exchange energy density for the H atom, and $b$ is an integer which 
is chosen to be 4. $\tilde{\epsilon}=-\epsilon_x^{\rm ex,conv}$ is the 
exact exchange energy density in the conventional gauge. This gauge 
function is integrated to zero, i.e., $\int d^3r~G(\rv) = 0$, as required. 
It satisfies the correct uniform coordinate scaling relation, 
$G_\lambda(\rv) = \lambda^4 G(\lambda\rv)$,  and non-uniform scaling 
relation $G_\lambda^x(x,y,z) = \lambda G(\lambda x,y,z)$.

\begin{figure}
\begin{center}
\includegraphics[width=\columnwidth]{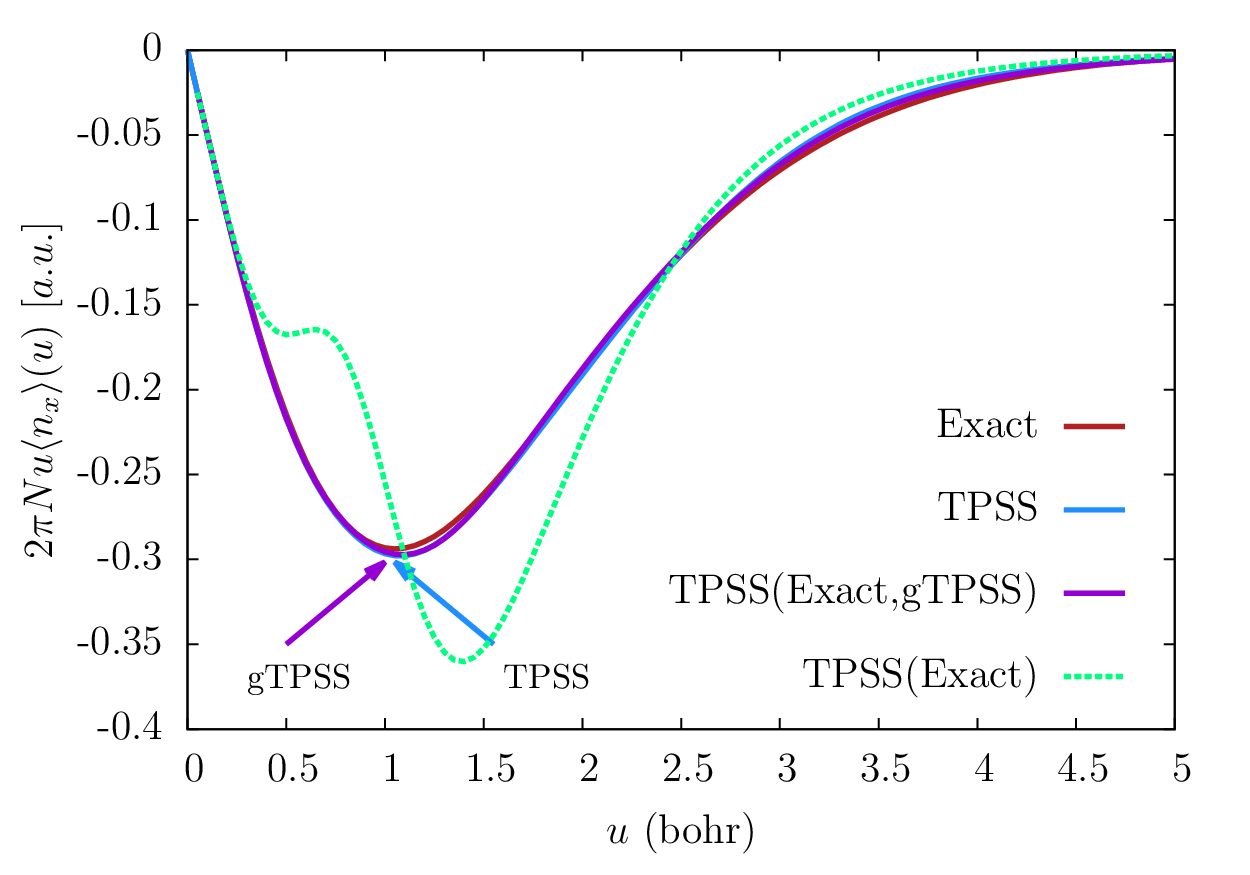}
\caption{Comparison of the system-average holes for the two-electron 
exponential density. ``Exact'' represents the conventional exact 
system-averaged hole, ``TPSS'' represents the present TPSS
system-averaged hole, ``Exact-gTPSS'' represents the system-averaged hole 
generated from the present gauge-corrected TPSS hole model with the 
conventional exact exchange energy density, and ``Exact-TPSS'' represents 
the exact system-averaged hole but it is calculated from the gauge-uncorrected 
TPSS hole model with the conventional exact exchange energy density.}
\label{figure8}
\end{center}
\end{figure}

However, in the density tail ($r\to\infty$) limit of an atom, the exact 
exchange energy density in the conventional gauge decays as 
$e_{\rm x}^{\rm ex,conv} \sim -n/2r$, but the TSSP gauge function decays 
as $G(\rv) \sim n$. As a result, the exchange energy density in this gauge 
becomes positive. In order to fix this deficiency, we impose a constraint 
on the density tail,  
\begin{eqnarray}\label{conv-negativity}
\lim_{ r \to \infty} \frac{G}{e_{\rm x}^{\rm conv}} = 0.
\end{eqnarray}
This can be achieved by requiring that in the $r\to\infty$ limit, $G$ 
decays as $n^p$ with $p > 1$. Here we choose $p=\frac{3}{2}$ and take 
the same form of the TSSP gauge function, but with $g$ abd $f$ given by
\begin{eqnarray}\label{newgauge}
g &=& \nabla \tilde{\epsilon}, \\ 
\label{ffunction-new}
f &=& \frac{ \Big( n/\tilde{\epsilon}^\frac{7}{3} \Big)^{3/2} }
{ 1 + c\Big(n/{\tilde{\epsilon}^3}\Big)^{5/2} }
\bigg(\frac{\tau^W}{\tau}\bigg)^b.
\end{eqnarray}
Here $a = 0.01799$ and $c = 0.00494$ are determined by fitting the  
system-averaged hole generated from the present TPSS shape function 
with the exact enhancement factor in the conventional gauge to the exact
system-averaged hole of the two-electron exponential density. The 
fitting procedure is the same as that in the determination of the
$H(s,z=1)$ function. The parameter $b=4$ remains the same as that in the
original version of Eq.~(\ref{ffunction}).

Figure~\ref{figure8} shows the comparison of the present TPSS
system-averaged exchange hole and the system-averaged exchange hole 
calculated from the present TPSS hole model with the conventional exact 
exchange energy density with and without gauge correction of 
Eqs.~(\ref{newgauge}) and~(\ref{ffunction-new} to the exact one. 
From Fig.~\ref{figure8} we can observe that the exact 
system-averaged exchange hole generated from the present TPSS hole model 
simply by replacing the TPSS enhancement factor with the conventional 
exact exchange enhancement factor without gauge correction
significantly deviates from the exact system-averaged hole. However, 
the agreement can be significantly improved with our gauge correction of 
Eqs.~(\ref{newgauge}) and~(\ref{ffunction-new}.

\begin{figure}
\begin{center}
\includegraphics[width=\columnwidth]{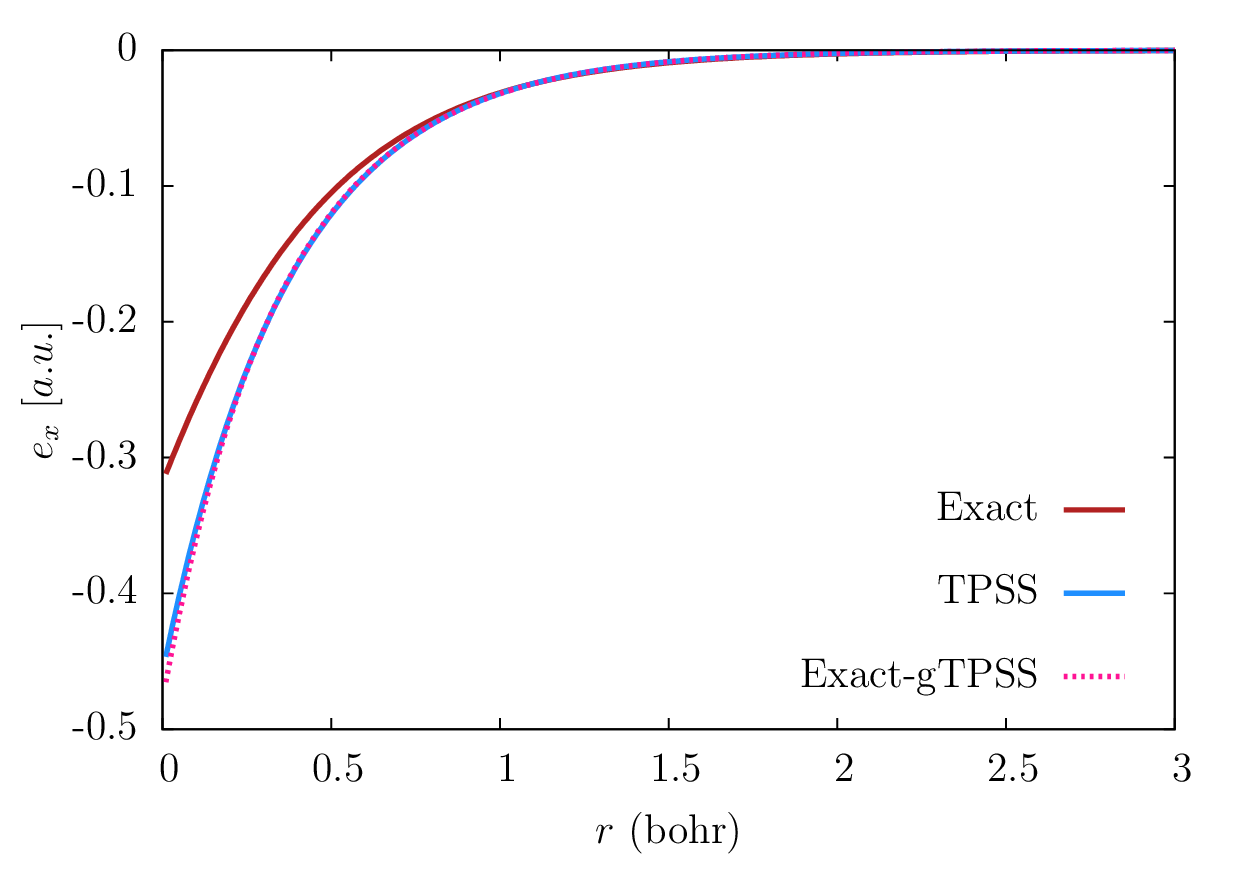}
\caption{Comparison of the exchange energy densities for the two-electron 
exponential density calculated with different approximations to the exact
one. ``TPSS'' represents the TPSS exchange energy density calculated 
directly from the TPSS exchange energy functional. ``gTPSS`` 
represents the gauge-corrected TPSS exchange energy density. }
\label{figure9}
\end{center}
\end{figure}

Figure~\ref{figure9} shows the comparison of the TPSS exchange 
energy density evaluated with the TPSS functional without and with 
gauge correction to the exact conventional exchange energy density
for the two-electron exponential density. From Fig.~\ref{figure9}, 
we can observe that gauge correction is important, leading to a 
better agreement of the gauge-corrected TPSS exchange energy density 
with the exact one. 

\section{Appication to range-separation density functional}
As a simple application, we apply the present TPSS hole model to construct
a range-separation functional. In general, there are two ways to construct 
range-separation functionals, depending on the need. For example, we may 
employ a semilocal DFT for the long-range part, while the exact exchange 
is used for the short-range part, as pioneered by Heyd, Scuseria and 
Ernzerhof~\cite{HSE03}. This kind of range-separation functional is 
developed for solids, particularly useful for metallic solids, because 
usual hybrids requires much larger momentum cutoff for metallic systems 
with electrons nonlocalized. We may employ a
semilocal DFT for the short-range part, while the exact exchange is used 
for the long-range part, as developed by Henderson 
{\em et al.}~\cite{Henderson08} on the basis of the PBE hole. This kind 
of range-separation functionals are usually developed for molecular 
calculations, because the improved long-range part of the exchange hole 
greatly benefits the description of molecular properties, without 
significant increase of computational cost. Many range-separation 
functionals have been proposed~\cite{hse06,Henderson07,AVKrukau08,
Henderson09,GalliPRB14,Truhlar11}. In the following, we will explore the 
TPSS hole-based range separation functional with TPSS exchange functional 
being the long-range (LR) part and the exact exchange being the short-range 
(SR) part, aiming to improve the TPSS description of two-small band gaps 
and reaction barrier heights.      

The idea of the construction of our TPSS-based range-separation functional 
is rooted in the constrcution of usual one-parameter hybrid functionals, 
which, in general, can be written as
\begin{eqnarray}
E_{xc}^{\rm hybrid} = aE_{\rm x}^{\rm HF} + (1-a)E_{\rm x}^{\rm sl} 
+ E_{\rm c}^{\rm sl}, 
\end{eqnarray}
where $a$ is the mixing parameter that controls the amount of exact 
exchange mixed into a semilocal (sl) functional. 

\begin{table}[t]
\caption{Band gaps (in eV) calculated with LSDA, PBE, TPSS, HSE and
TPSS-based range-separation functional with $a=0.25$ and $\omega=0.10$
(PW = present work), compared to experiments.}
\begin{tabular}{l|ccccccc}
\hline
\hline
     &LSDA   &PBE     & TPSS   &HSE    & PW      & Exp \\
C    &4.17   &4.2     &4.24    &5.43   &5.48     &5.48 \\
CdSe &0.31   &0.63    &0.85    &1.48   &1.82     &1.90 \\
GaAs &0.04   &0.36    &0.6     &1.11   &1.44     &1.52 \\
GaN  &2.15   &2.22    &2.18    &3.48   &3.5      &3.50 \\
GaP  &1.56   &1.74    &1.83    &2.39   &2.53     &2.35 \\
Ge   &-      &0.13    &0.32    &0.8    &0.99     &0.74 \\
InAs &0      &-       &0.08    &0.57   &0.85     &0.41 \\
InN  &0      &0       &0       &0.72   &0.75     &0.69 \\
InSb &-      &0       &-       &0.47   &0.73     &0.23 \\
Si   &0.53   &0.62    &0.71    &1.2    &1.31     &1.17 \\
ZnS  &2.02   &2.3     &2.53    &3.44   &3.78     &3.66 \\
\hline
ME &$-.89$ &$-0.86$ &$-0.76$ &$-0.12$&0.14     & \\
\hline
MAE&0.89   &0.86    &0.76    &0.15   &0.17     & \\
\hline\hline
\end{tabular}
\label{table3}
\end{table}

\begin{table}[t]
\caption{AE6 atomization energies (in kcal/mol) calculated with LSDA, PBE,
TPSS, HSE and
TPSS-based range-separation functional with $a=0.25$ and $\omega=0.10$
(PW = present work), compared to experimental values~\cite{PCCP04}.}
\begin{tabular}{l|cccccccc}
\hline
\hline
                &LSDA   &PBE    & TPSS   &TPSSh    &HSE     & PW    & Expt \\
SiH$_4$         &347.4  &313.2  &333.7   &333.6    &314.5   &333.6  & 322.4\\
SiO             &223.9  &195.7  &186.7   &182.0    &182.1   &175.4  & 192.1\\
S$_2$           &135.1  &114.8  &108.7   &105.9    &106.3   &101.9  & 101.7\\
C$_3$H$_4$      &802.1  &721.2  &707.5   &704.4    &705.9   &699.9  & 704.8\\
C$_2$H$_2$O$_2$ &754.9  &665.1  &636.0   &628.0    &635.3   &616.4  & 633.4\\
C$_4$H$_8$      &1304   &1168   &1156    &1154     &1152    &1152   & 1149 \\
\hline
ME              &77.4   &12.4   &4.1     &0.75     &$-1.2$  &$-4.0$ & \\
\hline
MAE             &77.4   &15.5   &5.9     &6.1      &4.8     &8.8    & \\
\hline\hline
\end{tabular}
\label{table4}
\end{table}

Following the prescription of Heyd, Scuseria and Ernzerhof 
(HSE)~\cite{HSE03}, we write the TPSS-based range-separation functional as 
\begin{eqnarray}\label{rsfunctional}
E_{\rm xc} = aE_{\rm x}^{\rm HF,SR} + (1-a)E_{\rm x}^{\rm sl,SR} 
+ E_{\rm x}^{\rm sl,LR} + E_{\rm c}^{\rm sl}, 
\end{eqnarray}
where $E_{\rm x}^{\rm HF,SR}$ is the Hartree-Fock (HF) exchange serving as 
part of the short-range contribution, while $E_{\rm x}^{\rm sl,SR}$ is the 
TPSS exchange that provides the rest of the short-range contribution. 
$E_{\rm c}^{\rm sl}$ is TPSS correlation. The long-range contribution is 
provided fully by the TPSS exchange $E_{\rm x}^{\rm sl,LR}$. They are given, 
respectively, by
\begin{eqnarray}\label{hfsr}
\epsilon_{\rm X}^{{\rm HF,SR}} = 
\frac{1}{2}\int_0^{\infty}du~4\pi u^2 h_{\rm X}^{\rm HF}(\rv,u) 
\frac{{\rm erfc}(\omega u)}{u},
\end{eqnarray}
\begin{eqnarray}\label{slsr}
\epsilon_{\rm X}^{{\rm sl,SR}} = 
\frac{1}{2}\int_0^{\infty}du~4\pi u^2 h_{\rm X}^{\rm DFT}(\rv,u) 
\frac{{\rm erfc}(\omega u)}{u},
\end{eqnarray}
\begin{eqnarray}\label{sllr}
\epsilon_{\rm X}^{{\rm sl,LR}} = 
\frac{1}{2}\int_0^{\infty}du~4\pi u^2 h_{\rm X}^{\rm DFT}(\rv,u) 
\frac{{\rm erf}(\omega u)}{u},
\end{eqnarray}
where $\omega$ is a range-separation parameter, and ${\rm erf}(x)$ is the 
error function defined by Eq.~(\ref{errorf}).
From Eqs.~(\ref{rsfunctional})-(\ref{sllr}), we can see that the amount of
exact exchange mixing is controlled by two parameters $a$ and $\omega$. 
Determination of them is discussed below. To test this functional, we
have implemented it into the development version of 
Gaussian 09~\cite{gaussian09}. 

\begin{table*}[t]
\caption{BH6 reaction barrier heights (in kcal/mol) calculated with PBE,
TPSS, HSE and TPSS-based range-separation functional with $a=0.25$ and
$\omega=0.10$ (PW = present work), in comparison with reference
values~\cite{JPaier10,JPCA03}. f(r) = forward (reverse) barrier height.}
\begin{tabular}{l|cccccccc}
\hline
\hline
                                          &PBE       &TPSS     &TPSSh   &HSE     & PW       &Ref \\
OH + CH$_4$ $\rightarrow$ CH$_3$ + H$_2$O &$-5.29$   &$-0.97$  &1.50    &1.96    & 4.86     &6.54(f) \\
                                          &8.95      &9.90     &11.79   &13.9    & 14.3     &19.6(r) \\
\hline
H + OH $\rightarrow$ O + H$_2$            &3.69      &$-1.56$  &$-0.15$ &7.06    & 1.75     &10.5(f) \\
                                          &$-1.47$   &4.73     &6.90    &5.93    & 9.89     &12.9(r) \\
\hline
H + H$_2$S $\rightarrow$ H$_2$ + HS       &$-1.20$   &$-4.55$  &$-3.72$ &1.03    & $-2.64$  &3.55(f) \\
                                          &9.40      &12.72    &13.4    &12.4    & 14.4     &17.3(r) \\
\hline
ME                                        &$-9.37$   &$-8.34$  &$-6.76$ &$-4.66$ & $-4.63$  & \\
\hline
MAE                                       &9.37      &8.34     &6.76    &4.66    & 4.63     & \\
\hline\hline
\end{tabular}
\label{table5}
\end{table*}

In the TPSS-based hybrid functional TPSSh, $a=0.1$ was fitted to molecular
properties. In other words, in TPSSH, the optimized of $a$ is 0.1. So, 
in the TPSS-based range-separation functional of Eq.~(\ref{rsfunctional}),
the best value of $\omega$ will be 0, if $a=0.1$ is chosen. Since in the 
range-separation functional, some amount of the exact exchange (here the
long-range part) in TPSSh is replaced by semilocal TPSS functional, to
compensate for this, we need a value of $a$ larger than 0.1. Then we can
find the best range-separation parameter $\omega$ by fitting to some
electronic properties. To avoid possible overfitting, here we choose 
$a=1/4$, a value that was recommended by Perdew, Ernzerhof, and 
Burke~\cite{Perdew96} and adopted with PBE0 functional~\cite{PBE0}. The
parameter $\omega$ is determined by a fit to the band gap of diamond 
(C). This yields $\omega=0.1$. Then we apply this range-separation 
functional to calculate the band gaps of 10 semiconductors. The results 
are listed in Table~\ref{table3}. From Table~\ref{table3}, we see that the 
band gaps of this range-separation functional is remarkably accurate, 
with a mean 
absolute deviation from experiments of only $0.17$ eV, about the same 
accuracy of HSE functional. We can also see from Table~\ref{table3} that
this TPSS-based range-separation functional should provide more accurate 
description for large band-gap materials, and therefore it provides an
alternative choice for band-gap and other solid-state calculations.

Next, we apply our range-separation functional to calculate atomization 
energies of AE6 molecules. The results are listed in Tables~\ref{table4}. 
From Tables~\ref{table4}, we can see that, our range-separation 
functional only worsens the TPSS atomization energies of TPSS functional
for this special set by about 3 kcal/mol. This error is still smaller 
than many other DFT methods such as LSDA, PBE, and PBEsol. 

Reaction barrier heights are decisive quantity in the study of chemical
kinetics. However, semilocal functionals tend to underestimate this 
quantity. As an interesting application, we apply our range-separation 
functional to calculate representative reaction barrier heights of BH6, 
which consists of three forward (f) and three reverse (r) barrier heights. 
The results are listed in Table~\ref{table5}. For comparison, we also
calculated these barrier heights using PBE, TPSS, TPSSh, and HSE. From
Table~\ref{table5}, we obserce that our range-separation functional
provides the most accurate description of BH6, compared to other 
functionals considered.

\section{Conclusion}
In conclusion, we have developed a semilocal exchange hole underlying TPSS
exchange functional. The hole is exact in the uniform-gas limit and 
accurate for iso-orital densities. Then is is interpolated between these
two densities through WVA on the surface of infinite barrier model. Our 
numerical test on the Ne atom shows that the hole mimics the conventional
exact exchange hole quite accurately. In particular, with our new gauge
function correction, the hole model can generate the system-averaged hole
from the conventional exact exchange energy density in good agreement
with the present TPSS hole and the exact hole for the two-electron 
exponential density. 

Based on the TPSS hole, we have constructed a range-separation functional.
This functional yields very accurate band gaps and reaction barrier heights,
without losing much accuracy on atomization energies. This accuracy makes
this functional an attractive alternative in solid-state calculations and
molecular kinetics.


\begin{thebibliography}{1}
\bibitem{ks65}
W. Kohn and L.J. Sham, Phys. Rev. {\bf 140}, A1133 (1965).
\bibitem{Parr89}
R.G. Parr and W. Yang, {\em Density Functional Theory of Atoms and Molecules}
(Oxford University Press, New York, 1989).

\bibitem{RMDreizler90}
R.M. Dreizler and E.K.U. Gross,
{\em Density Functional Theory} (Springer, Berlin, 1990).

\bibitem{PW86}
J.P. Perdew and Y. Wang, Phys. Rev. B {\bf 33}, 8800 (1986).

\bibitem{B88}
A.D. Becke, Phys. Rev. A {\bf 38}, 3098 (1988).

\bibitem{LYP88}
C. Lee, W. Yang, and R.G. Parr,
Phys. Rev. B {\bf 37}, 785 (1988).


\bibitem{BR89}
A.D. Becke and M. R. Roussel, Phys. Rev. A {\bf 39}, 3761 (1989).


\bibitem{B3PW91}
A.D. Becke,
J. Chem. Phys. {\bf 104}, 1040 (1996).

\bibitem{B3LYP}
P.J. Stephens, F.J. Devlin, C.F. Chabalowski, and M.J. Frisch,
J. Phys. Chem. {\bf 98}, 11623 (1994).



\bibitem{PBE96}
J.P. Perdew, K. Burke, and M. Ernzerhof, Phys. Rev. Lett. {\bf 77},
3865 (1996).

\bibitem{VSXC98}
T.V. Voorhis and G.E. Scuseria, J. Chem. Phys. {\bf 109}, 400 (1998).


\bibitem{HCTH}
F.A. Hamprecht, A.J. Cohen, D.J. Tozer, and N.C. Handy,
J. Chem. Phys. {\bf 109}, 6264 (1998).

\bibitem{PBE0}
M. Ernzerhof and G.E. Scuseria,
J. Chem. Phys. {\bf 110}, 5029 (1999).

\bibitem{HSE03}
J. Heyd, G.E. Scuseria, and M. Ernzerhof,
J. Chem. Phys. {\bf 118}, 8207 (2003).

\bibitem{AE05}
R. Armiento and A.E. Mattsson,
Phys. Rev. B {\bf 72}, 085108 (2005).


\bibitem{MO6L}
Y. Zhao and D.G. Truhlar,
J. Chem. Phys. {\bf 125}, 194101 (2006).

\bibitem{TPSS03}
J. Tao, J.P. Perdew, V.N. Staroverov, and G.E. Scuseria,
Phys. Rev. Lett. {\bf 91}, 146401 (2003).


\bibitem{revTPSS}
J.P. Perdew, A. Ruzsinszky, G.I. Csonka, L.A. Constantin, and J. Sun,
Phys. Rev. Lett. {\bf 103}, 026403 (2009).

\bibitem{PBEsol}
J.P. Perdew, A. Ruzsinszky, G.I. Csonka, O.A. Vydrov, G.E.
Scuseria, L.A. Constantin, X. Zhou, and K. Burke, 
Phys. Rev. Lett. {\bf 100}, 136406 (2008).

\bibitem{SCAN15}
J. Sun, A. Ruzsinszky, and J.P. Perdew,
Phys. Rev. Lett. {\bf 115}, 036402 (2015).

\bibitem{Kaup14}
A.V. Arbuznikov and M. Kaupp,
J. Chem. Phys. {\bf 141}, 204101 (2014).

\bibitem{Tao-Mo16}
J. Tao and Y. Mo,
Phys. Rev. Lett. {\bf 117}, 073001 (2016).

\bibitem{VWN80}
S.H. Vosko, L. Wilk, and M. Nusair,
Can. J. Phys. {\bf 58}, 1200 (1980).

\bibitem{PW92}
J.P. Perdew and Y. Wang,
Phys. Rev. B {\bf 45}, 13244 (1992).

\bibitem{PW91}
J.P. Perdew, J.A. Chevary, S.H. Vosko, K.A. Jackson, M.R. Pederson,
D.J. Singh, and C. Fiolhais,
Phys. Rev. B {\bf 46}, 6671 (1992).

\bibitem{ZWu06}
Z. Wu and R.E. Cohen,
Phys. Rev. B {\bf 73}, 235116 (2006).


\bibitem{CJCramer09}
C.J. Cramer and D.G. Truhlar, 
Phys.Chem.Chem.Phys. {\bf 11}, 10757 (2009).

\bibitem{Quest12}
R. Peverati and D.G. Truhlar,
Phys. Phil.Trans. R. Soc. A {\bf 372}, 20120476 (2011). 


\bibitem{Yangreview}
A.J. Cohen, P. Mori-Sanchez, and W. Yang,
Chem.Rev. {\bf 112}, 289 (2012).

\bibitem{Becke14}
A.D. Becke,
J. Chem. Phys. {\bf 140}, 18A301 (2014).

\bibitem{VNS03}
V.N. Staroverov, G.E. Scuseria, J. Tao, and J.P. Perdew,
J. Chem. Phys. {\bf 119}, 12129 (2003); 121, 11507(E) (2004).

\bibitem{PHao13}
P. Hao, J. Sun, B. Xiao, A. Ruzsinszky, G.I. Csonka, J. Tao, S. Glindmeyer, 
and J.P. Perdew, 
J. Chem. Theory Comput. {\bf 9}, 355 (2013).

\bibitem{LGoerigk10}
L. Goerigk and S. Grimme,
J. Chem. Theory Comput. {\bf 6}, 107 (2010).

\bibitem{LGoerigk11}
L. Goerigk and S. Grimme, 
J. Chem. Theory Comput. {\bf 7}, 291 (2011).

\bibitem{CAdamo10}
D. Jacquemin and C. Adamo,
J. Chem. Theory Comput. {\bf 7}, 369 (2011).

\bibitem{YMo16}
Y. Mo, G. Tian, R. Car, V.N. Staroverov, G.E. Scuseria, and J. Tao,
submitted.


\bibitem{Csonka09}
G.I. Csonka, J.P. Perdew, A. Ruzsinszky, P.H.T. Philipsen, S. Leb\'egue, 
J. Paier, O.A. Vydrov, and J.G. \'Angy\'an
Phys. Rev. B {\bf 79}, 155107 (2009).

\bibitem{PBlaha09}
P. Haas, F. Tran, and P. Blaha,
Phys. Rev. B {\bf 79}, 085104 (2009).


\bibitem{FTran16}
F. Tran, J. Stelzl, and P. Blaha,
J. Chem. Phys. {\bf 144}, 204120 (2016).

\bibitem{VNS04}
V.N. Staroverov, G.E. Scuseria, J. Tao, and J.P. Perdew,
Phys. Rev. B {\bf 69}, 075102 (2004).

\bibitem{PSTS08}
J.P. Perdew, V.N. Staroverov, J. Tao, and G.E. Scuseria,
Phys. Rev. A {\bf 78}, 052513 (2008).


\bibitem{JJaramillo03}
J. Jaramillo, G.E. Scuseria, and M. Ernzerhof,
J. Chem. Phys. {\bf 118}, 1068 (2003).

\bibitem{Truhlar11}
R. Peverati and D.G. Truhlar,
Phys. Chem. Lett. {\bf 2}, 2810 (2011).

\bibitem{PBY96}
Phys. Rev. B {\bf 54}, 16533 (1996).

\bibitem{Taobook10}
J. Tao,
{\em Density Functional Theory of Atoms, Molecules, and Solids}
(VDM Verlag, Germany, 2010).


\bibitem{EP98}
 M. Ernzerhof and J.P. Perdew
J. Chem. Phys. {\bf 109}, 3313 (1998).

\bibitem{tpsshole}
L.A. Constantin, J.P. Perdew, and J. Tao,
Phys. Rev. B {\bf 73}, 205104 (2006).

\bibitem{Lucian13}
L.A. Constantin, E. Fabiano, and F. Della Sala
Phys. Rev. B {\bf 88}, 125112 (2013).

\bibitem{LKronik11}
Phys. Rev. B {\bf 84}, 075144 (2011).

\bibitem{RBaer10}
Ann. Rev. Phys. Chem. {\bf 61}, 85 (2010).

\bibitem{JCTC16}
M. Modrzejewski, M. Hapka, G. Chalasinski, and M.M. Szczesniak,
J. Chem. Theory Comput., {\bf 12}, 3662 (2016).

\bibitem{Kresse11}
L. Schimka, J. Harl, and G. Kresse,
J. Chem. Phys. {\bf 134}, 024116 (2011).


\bibitem{tsspg08}
J. Tao, J.P. Perdew, V.N. Staroverov, and G.E. Scuseria,
Phys. Rev. A {\bf 77}, 012509 (2008).


\bibitem{OP79}
G.L. Oliver and J.P. Perdew, Phys. Rev. A {\bf 20}, 397 (1979).

\bibitem{KBurke98}
K. Burke, F.G. Cruz, and K.-C. Lam,
J. Chem. Phys. {\bf 109}, 8161 (1998).

\bibitem{PKZB}
J.P. Perdew, S. Kurth, A. Zupan, and P. Blaha, Phys. Rev. Lett. {\bf 82},
2544 (1999).


\bibitem{TLZR15}
J. Tao, S. Liu, Z. Fan, and A.M. Rappe,
Phys. Rev. B {\bf 92}, 060401(R) (2015).

\bibitem{Henderson08}
T.M. Henderson, B.G. Janesko, and G.E. Scuseria,
J. Chem. Phys. {\bf 128}, 194105 (2008).



\bibitem{LMiglio81}
L. Miglio, M.P. Tosi, and N.H. March,
Sur. Sci. {\bf 111}, 119 (1981).


\bibitem{IDMoore76}
I.D. Moore and N.H. March, Ann. Phys. {\bf 97}, 136 (1976).

\bibitem{Langreth82}
D.C. Langreth and J.P. Perdew, 
Phys. Rev. B {\bf 26}, 2810 (1982).


\bibitem{Harris74}
J. Harris and R.O. Jones, 
J. Phys. F {\bf 4}, 1170 (1974).

\bibitem{CQMa79}
C.Q. Ma and V. Sahni, 
Phys. Rev. B {\bf 20}, 2291 (1979).


\bibitem{DSG01}
F. Della Sala and A. G\"orling,
J. Chem. Phys. {\bf 115}, 5718 (2001).













\bibitem{hse06}
A.V. Krukau, O.A. Vydrov, A.F. Izmaylov, and G.E. Scuseria,
J. Chem. Phys. {\bf 125}, 224106 (2006).

\bibitem{Henderson07}
T.M. Henderson, A.F. Izmaylov, G.E. Scuseria, and A. Savin,
J. Chem. Phys. {\bf 127}, 221103 (2007).

\bibitem{AVKrukau08}
A.V. Krukau, G.E. Scuseria, J.P. Perdew, and A. Savin,
J. Chem. Phys. {\bf 129}, 124103 (2008).

\bibitem{Henderson09}
T.M. Henderson, B.G. Janesko, G.E. Scuseria, and A. Savin,
Int. J. Quantum Chem. {\bf 109}, 2023 (2009).

\bibitem{GalliPRB14}
J.H. Skone, M. Govoni, and G. Galli,
Phys. Rev. B {\bf 89}, 195112 (2014).


\bibitem{PCCP04}
Y. Zhao, J. Pu, B.J. Lynch, and D.G. Truhlar, 
Phys. Chem. Chem. Phys. {\bf 6}, 673 (2004).

\bibitem{JPaier10}
J. Paier, B.G. Janesko, T.M. Henderson, G.E. Scuseria, A. G\"uneis, and 
G. Kresse,
J. Chem. Phys. {\bf 132}, 094103 (2010).

\bibitem{JPCA03}
B.J. Lynch and D.G. Truhlar,
J. Phys. Chem. A {\bf 107}, 8996 (2003).

\bibitem{gaussian09}
Gaussian Development Version, Revision B.01, M.J. Frisch {\it et al}.,
Gaussian, Inc., Wallingford CT (2009).
 

\bibitem{Perdew96}
J.P. Perdew, M. Ernzerhof, and K. Burke,
J. Chem. Phys. {\bf 105}, 9982 (1996).


\end{thebibliography}
\end{document}